\documentclass[final,1p,times]{elsarticle} 
\usepackage{graphicx} 
\usepackage{amssymb} 
\usepackage{amsthm} 
\usepackage{lineno} 

\usepackage{amsmath,amssymb}
\newcommand{\bm}[1]{\mbox{\boldmath$#1$}}

\journal{Nuclear Physics A} 
\begin{document} 

\begin{frontmatter} 


\title{Spectral Properties of Quarks at Finite Temperature in Lattice QCD}

\author{ Masakiyo Kitazawa$^{a}$ and Frithjof Karsch$^{b}$ }

\address[a]{Department of Physics, Osaka University, 
Toyonaka, Osaka, 560-0043 Japan }
\address[b]{Brookhaven National Laboratory,
Bldg.~510A, Upton 11973, USA}

\begin{abstract} 
We analyze the quark spectral function
above and below the critical temperature for deconfinement
and at finite momentum in quenched lattice QCD. 
It is found that the temporal quark correlation function
in the deconfined phase near the critical temperature 
is well reproduced by a two-pole ansatz for the spectral 
function.
The bare quark mass and momentum dependences of the 
spectral function are analyzed with this ansatz.
In the chiral limit we find that even near the critical 
temperature the quark spectral function has two collective 
modes corresponding to the normal and plasmino 
excitations in the high temperature ($T$) limit.
The pole mass of these modes at zero momentum, 
which should be identified to be the thermal mass of the quark, 
is approximately proportional to $T$ in a rather wide 
range of $T$ in the deconfined phase.
\end{abstract} 

\end{frontmatter} 



\section{Introduction}

After the advent of Relativistic Heavy Ion Collider (RHIC),
much attention has been paid for the properties of matter near 
and above the critical temperature of QCD phase transition, $T_c$.
To understand the structure of the matter in this region, 
it is desirable to identify the basic degrees of freedom of 
the system and their excitation properties.
In the present study, we analyze the spectral properties of quarks,
which are one of the fundamental degrees of freedom of QCD,
at finite temperature above and below $T_c$ using quenched 
lattice QCD with clover improved fermions \cite{KK1,KK2}.

At asymptotically high temperatures, one can calculate the 
quark propagator using perturbative techniques.
It is known that the collective excitations of quarks in this 
limit develop a mass gap (thermal mass) that is proportional to $gT$,
where $g$ and $T$ denote the gauge coupling and temperature, 
respectively \cite{plasmino}. 
Moreover, in this limit the number of poles in the quark propagator 
is doubled; in addition to the normal modes, which reduce to poles 
in the free particle propagator, plasmino modes appear.

In order to see the properties of quark spectral function
$\rho( \omega,\bm{p} )$ in this limit, let us first consider
the spectral function at zero momentum. In this case,
the Dirac structure of $\rho( \omega,\bm{p} )$ can be decomposed 
using the projection operators
$L_\pm = ( 1 \pm \gamma^0 )/2$ as 
\begin{align}
\rho( \omega, \bm{0} )
= \rho^{\rm M}_+( \omega ) L_+ \gamma^0 
+ \rho^{\rm M}_-( \omega ) L_- \gamma^0 .
\label{eq:rho^M}
\end{align}
While $\rho^{\rm M}_\pm(\omega)$ for free quarks are given by
$ \rho_\pm^{\rm M} ( \omega ) = \delta( \omega \mp m ) $,
in the high temperature limit one obtains 
$ \rho_\pm^{\rm M} ( \omega ) = 
[ \delta( \omega - m_T ) + \delta( \omega + m_T ) ] /2 $,
where $m_T=gT/\sqrt6$ is the thermal mass.
The two poles at $\omega=\pm m_T$ correspond to the 
normal and plasmino modes.

The Dirac structure of $\rho( \omega,\bm{p} )$ 
in the high temperature limit is also decomposed as
\begin{align}
\rho( \omega,\bm{p} )
= \rho^{\rm P}_+( \omega,p ) P_+(\bm{p}) \gamma^0
+ \rho^{\rm P}_-( \omega,p ) P_-(\bm{p}) \gamma^0 ,
\label{eq:rho^P}
\end{align}
with the projection operators
$P_\pm (\bm{p})= ( 1 \pm \gamma^0\hat{\bm{p}}\cdot\bm{\gamma}) )/2$
and $p=|\bm{p}|$.
The spectral functions $\rho^{\rm P}_\pm(\omega,\bm{p})$ 
in the high temperature limit read
\begin{align}
\rho_\pm^{\rm P} ( \omega,p )
= Z_1(p) \delta( \omega \mp E_1(p) )
+ Z_2(p) \delta( \omega \pm E_2(p) )
+ \rho_{\rm cont.}( \pm\omega,p ),
\label{eq:rho^P_HTL}
\end{align}
where $\rho_{\rm cont.}( \omega,p )$ represents the 
contribution of the continuum
taking non-zero values in the space-like region, and
poles at $E_1(p)>0$ and $E_2(p)>0$ corresponds to the normal
and plasmino modes, respectively.
The dispersion relation of the plasmino has a minimum at 
nonzero $p$ \cite{plasmino}.

\section{Quark spectral function at zero momentum}

In this section, we analyze the quark spectral function 
for zero momentum but with finite bare quark mass on the lattice.
We use gauge field ensembles which have been generated and 
used previously by the Bielefeld group to study screening 
masses and spectral functions \cite{Bielefeld,KK2}.
The quark spectral function in this case is decomposed into 
$\rho^{\rm M}_\pm(\omega)$ as in Eq.~(\ref{eq:rho^M}).
In order to extract the spectral function 
$\rho^{\rm M}_+(\omega)$ from the lattice correlation 
function, we assume a simple ansatz for the shape of 
$\rho^{\rm M}_+(\omega)$ including few fitting parameters.
We found that the two-pole ansatz for $\rho^{\rm M}_+(\omega)$,
\begin{align}
\rho^{\rm M}_+(\omega)
= Z_1 \delta( \omega - E_1 ) + Z_2 \delta( \omega + E_2 ),
\label{eq:2pole}
\end{align}
reproduces the lattice correlator quite well,
where $Z_{1,2}$, and $E_{1,2}>0$ are fitting parameters
determined from the correlated fit.
The pole at $\omega=-E_2$ in Eq.~(\ref{eq:2pole}) 
corresponds to the plasmino mode.
The success of two-pole ansatz for the quark correlation 
functions suggests that the positions of the poles of 
the quark propagator would be near the real axis at 
$\omega=E_1$ and $-E_2$ with small imaginary parts.
Provided that the positions of poles of the propagator are 
gauge independent, this also indicates that 
our results on the fitting parameters $E_1$ and $E_2$ have 
small gauge dependence.

\begin{figure}[th]
\centering
\includegraphics[scale=1]{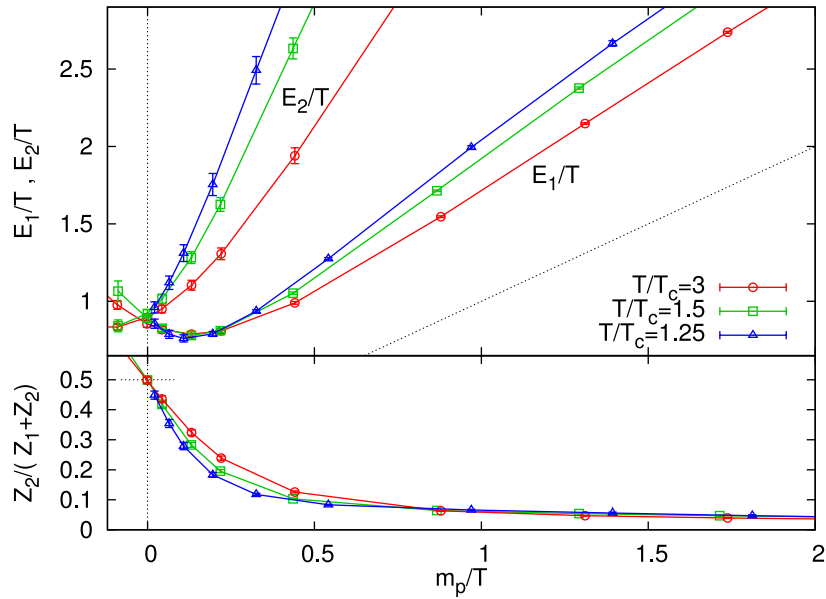}	       
\caption[]{
Bare quark mass dependence of fitting parameters $E_{1,2}$ 
and the relative strength of the plasmino mode, 
$Z_2 / ( Z_1+Z_2 )$, at $T/T_c=1.25$, $1.5$ and $3$ obtained 
from calculations on lattice of size $64^3\times16$.
}
\label{fig:ZE}
\end{figure}

In Fig.~\ref{fig:ZE}, we show the dependence of 
$E_1$, $E_2$ and $Z_2 / ( Z_1+Z_2 )$ on the bare quark
mass, $m_p$, for $T/T_c=1.25$, $1.5$, and $3$
obtained from calculations on 
lattices of size $64^3\times16$.
The figure shows that the ratio $Z_2 / (Z_1+Z_2)$ 
becomes larger with decreasing $m_p$ and eventually 
reaches $0.5$ irrespective of $T$.
The numerical result for each $T$ shows that $E_1=E_2$ 
is satisfied within statistical errors there.
These results mean that the quark propagator is chirally 
symmetric at this point \cite{KK2}.
Moreover, $\rho^{\rm M}_+(\omega)$ at this point 
has the same form as the spectral function in the high 
temperature limit.
We therefore define the thermal mass of the quark on the 
lattice as $m_T \equiv ( E_1+E_2 )/2$ at this point.
One finds that the ratio $m_T/T$ is insensitive 
to $T$ in the range analyzed in this work, 
although it becomes slightly larger with decreasing $T$,
which would be in accordance with the expected parametric
form at high temperature, $m_T\sim gT$.

Figure~\ref{fig:ZE} also shows that the relative strength 
of the plasmino pole,
$Z_2/(Z_1+Z_2)$, decreases with increasing values
of the bare mass, $m_p$. The spectral function
$\rho^{\rm M}_+(\omega)$ thus will
eventually be dominated by a single-pole.
This result agrees with the perturbative result in Yukawa
models that $\rho^{\rm M}_+(\omega)$ approaches the spectral 
function of free quarks as the bare quarks mass becomes larger
\cite{BBS92,KKN06}.

For $T<T_c$, we found that the lattice correlator is concave
in the log-scale plot.
This behavior indicates that the positivity condition of 
$\rho^{\rm M}_+(\omega)$ is violated below $T_c$.
In fact, we have checked that the two-pole ansatz 
Eq.~(\ref{eq:2pole}) gives unacceptably large $\chi^2/\rm{dof}$
below $T_c$.
It is also found that the quark correlator does not approach
the chirally symmetric one even in the chiral limit, in 
contrast to the previous result for $T>T_c$.
These results for $T<T_c$ seem consistent with a na\"ive picture
that quark excitations are confined and the chiral symmetry
is spontaneously broken below $T_c$.


\section{Quark spectral function at finite momentum}

\begin{figure}[th]
\centering
\includegraphics[scale=1]{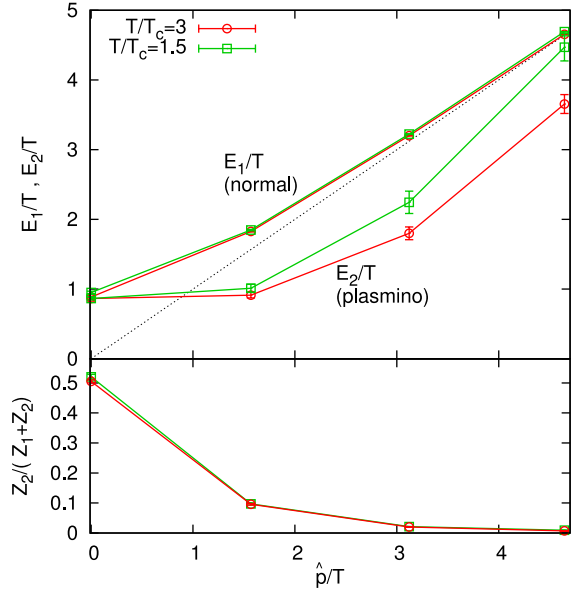}
\caption[]{
Dependences of the fitting parameters 
$E_1$ and $E_2$ and the ratio $Z_2/(Z_1+Z_2)$ on
the lattice momentum $\hat{p}=(1/a) \sin(pa)$
for $T/T_c=1.5$ and $3$.
}
\label{fig:p>0}
\end{figure}

Next let us analyze the quark spectral function 
at finite momentum on lattices with size $64^3\times16$
for $T/T_c=1.5$ and $3$.
Throughout this section we consider the quark propagator 
in the chiral limit.
The quark propagator in the chiral limit is 
decomposed into $\rho^{\rm P}_\pm(\omega,p)$
according to Eq.~(\ref{eq:rho^P}).
Following the same approach used in the previous section,
we adopt the two-pole ansatz 
\begin{align}
\rho_+^{\rm P}(\omega,p)
= Z_1 \delta( \omega - E_1 ) + Z_2 \delta( \omega + E_2 ) ,
\label{eq:2pole_p}
\end{align}
and determine four parameters from a correlated fit.
The $\delta$-functions at $\omega=E_1$ and $-E_2$ correspond
to the normal and plasmino modes, respectively.
We found that $\chi^2/{\rm dof}$ with this ansatz is always 
smaller than $1.5$ for all momenta analyzed in this study.
This result means that the two-pole ansatz again reproduces 
the lattice correlation function well.

In Fig.~\ref{fig:p>0}, we show the momentum dependence
of the fitting parameters $E_1$, $E_2$, and 
$Z_2/(Z_1+Z_2)$ for $T/T_c=1.5$ and $3$.
The horizontal axis represents the momentum 
on the lattice $\hat{p}=(1/a) \sin pa$.
The figure shows that for large momentum 
$Z_2/(Z_1+Z_2)$ rapidly decreases and $E_1$ approaches 
the light cone.
The spectral function at large momentum therefore 
approaches that of a free quark, consistent with
the perturbative result.
One also finds that $E_2$ is always smaller than $E_1$,
in contrast to the results in the previous section.
This behavior qualitatively agrees with the behavior of
poles in the high $T$ limit \cite{plasmino}.
One also observes from Fig.~\ref{fig:p>0} that $E_2$ enters
the space-like region at high momentum.

An interesting property of the quark propagator in the 
high temperature limit Eq.~(\ref{eq:rho^P_HTL}) is that the 
dispersion relation
of the plasmino has a minimum at finite momentum.
In Fig.~\ref{fig:p>0}, one sees that the value of $E_2$ 
at lowest non-zero momentum on our lattice, 
$ p_{\rm min} = 2\pi T(N_\tau/N_\sigma) \simeq 1.5T $,
is slightly larger than that at zero momentum,
and the existence of such a minimum is 
suggested but not yet confirmed in the present analysis.

In summary, we analyzed the dependence of the quark 
spectral function on temperature $T$, bare quark mass $m$, and 
momentum $p$ in quenched lattice QCD with Landau gauge fixing.
Above $T_c$, we found that the two-pole approximations for 
the spectral functions in the projected channels, 
$\rho^{\rm M}_\pm(\omega)$ and $\rho^{\rm P}_\pm(\omega,p)$, 
can well reproduce the lattice correlation functions.
Although further studies on the volume dependence is
needed, this result indicates that the excitations of
quarks have small decay width even near $T_c$.
Below $T_c$, on the other hand, the two-pole ansatz fails 
completely.

The lattice simulations presented in this work have been 
carried out using the cluster computers ARMINIUS@Paderborn, 
BEN@ECT* and BAM@Bielefeld.

\end{document}